\documentclass{article}
\usepackage[utf8]{inputenc}
\usepackage{authblk}
\usepackage{setspace}
\usepackage[margin=1.25in]{geometry}
\usepackage{graphicx}
\graphicspath{ {./figures/} }
\usepackage{subcaption}
\usepackage{amsmath}
\usepackage{lineno}
\usepackage{amsmath}
\usepackage{graphicx}
\usepackage{dcolumn}
\usepackage{xcolor}
\usepackage{cite}
\usepackage{mathrsfs}
\usepackage{amssymb}
\usepackage{amsfonts}
\usepackage{bm}

\numberwithin{equation}{section}

	\title{Topological Transitions in a Kerr Nonlinear Oscillator}
	
	\author[1]{Juan Lin}
	\author[1]{Shou-Bang Yang}
	\author[1*]{Fan Wu}
	\author[1$\dag$]{Zhen-Biao Yang}
	\affil[1]{Fujian Key Laboratory of Quantum Information and Quantum Optics, College of Physics and Information Engineering, Fuzhou University, Fuzhou, Fujian 350116, China}
	\affil[*]{Address correspondence to:t21060@fzu.edu.cn}
	\affil[$\dag$]{Address correspondence to:zbyang@fzu.edu.cn}
	\date{\today }
	
\begin{document}
	\maketitle
\begin{abstract}
A Kerr nonlinear oscillator (KNO) supports a pair of steady eigenstates, coherent states with opposite phases, that are good for the encoding of continuous variable qubit basis states. Arbitrary control of the KNO confined within the steady state subspace allows extraction of the Berry curvature through the linear response of the physical observable to the quench velocity of the system, providing an effective method for the characterization of topology in the KNO. As an alternative, the control adopting the ``shortcut to adiabaticity" to the KNO enables the exploration of the topology through accelerated adiabatic eigenstate evolution to measure all three physical observables. Topological transitions are revealed by the jump of the first Chern number, obtained respectively from the integral of the Berry curvature and of the new polar angle relation, over the whole parameter space. Our strategy paves the way for measuring topological transitions in continuous variable systems.            
\end{abstract}


\section*{Introduction}  
Topology, with its abstract mathematical constructs, is concerned with the properties of geometric objects that are maintained under continuous deformation \cite{Schirber2016physics}. The light it casts upon the internal physics of geometric objects is of essential importance to the understanding of natural phenomena \cite{Klitzing2017Quantum}. Markedly, the concept of phases of matter has changed greatly due to the discovery of topological phases in condensed-matter systems \cite{Klitzing2017Quantum,Klitzing1980New,Moore2010The,Tsui1982Two,B2006Quantum,M2010Colloquium}. Topological phases can be characterized by topological invariants, such as the Chern number \cite{Chern1946Characteristic,Kohmoto1985Topological,Chiu2016Class}, which is of benefit to classify physical phenomena \cite{Chiu2016Class}. The characteristic of topological invariants remaining unchanged by small perturbations to the systems means that topological systems become potentially important tools of metrology \cite{Arai2018NatureCom} and powerful candidates for future quantum information processing platforms \cite{Campbell2017Roads, Melnikov2018Towards}.      

The past decade has witnessed great process for the direct experimental probing of topological phases in a wide variety of systems \cite{Shankar2022NRP}. Moreover, construction of versatile physical systems to investigate topological properties of various fundamental matter models is useful for exploration of quantum phenomena that are perhaps temporarily inaccessible \cite{Roushan2014Observation}. For instance, the basic topological conception implicated in the Haldane model has been focused and explored, through mapping the momentum space of such a condensed-matter model to the parameter space of a quantum system Hamiltonian \cite{Gritsev2012Dynamical}.  

The first Chern number $C_1$ \cite{Chern1946Characteristic}, as one of the basic topological invariants, is an integer. The jump between discrete values of the $C_1$  reveals nontrivial topological transitions between different topologically ordered phases \cite{Wen2004Quantum,Bernevig2013Topological}. For a quantum system whose Hamiltonian is written in terms of a set of externally controlled parameters, the $C_1$ is defined by integrating the Berry curvature $\textbf{B}$ over the closed manifold $\textbf{S}$ in the parameter space of the Hamiltonian $C_1=\frac{1}{2\pi}\oint_{S}\textbf{B}\cdot d\textbf{S}$ \cite{Berry1984Quantal}. This Chern number can be understood as counting the number of times an eigenstate wraps around a manifold in Hilbert space \cite{Hatsugai1993Chern,Niu1985Quantized}. From another perspective, the degenerate points in the parameter space of the Hamiltonian act as the sources or sinks of the Berry curvature $\textbf{B}$ (an effective magnetic field) and are analogs of monopoles \cite{Ville2009Creation,Wilczek1989Geometric,Zhang2016Quantum,Zhang2017Quantum}. From the viewpoint of Gauss' law, the $C_1$ represents the number of degenerate points enclosed by the parameter manifold $\textbf{S}$, and the topological transition of the $C_1$ occurs at the moment when the monopole passes in or out of the manifold $\textbf{S}$ \cite{Roushan2014Observation}.

Research has been done to structure the controllable Hamiltonian parameter manifold of, for instance, a microwave-addressed superconducting qubit, for measuring topological transitions \cite{Roushan2014Observation,Schroer2014Measuring,Tan2019PRL}. This is realized by obtaining the global topology characterized by the first Chern number, which is obtained through the integral of the Berry curvature extracted by the real-time observation of a certain physical observable \cite{Gritsev2012Dynamical,Roushan2014Observation,Schroer2014Measuring}. Such research has been generalized to interactions between two- or multiqubit systems \cite{Roushan2014Observation,Luo2016PRA,Lee2023npjquantuminformation}.

All such previous studies for the measurement of the topological invariant first Chern number $C_1$ are based on finite-dimensional basis of spin states \cite{Roushan2014Observation,Schroer2014Measuring,Tan2019PRL,Luo2016PRA,Lee2023npjquantuminformation}. There have been few investigations to observe topological transition induced by the discrete jumps of the $C_1$ with harmonic oscillator continuous variable states, for instance, coherent states \cite{Klauder1985APMP, Zhang1990RMP}, which can be spanned by infinite-dimensional Hilbert space formed with Fock states. In this work, we propose a method for the measurement of the first Chern number with a driven Kerr nonlinear oscillator (KNO), whose dynamics is controlled within the continuous variable subspace composed of a pair of coherent states with opposite phases.
 
Notice that in all such studies \cite{Roushan2014Observation,Schroer2014Measuring,Luo2016PRA,Lee2023npjquantuminformation}, the Berry curvature is dependent on the linear response of the physical observable to the slow changing rate (also referred to as the `quench velocity') of the control parameter \cite{Gritsev2012Dynamical}, which means it suffers from a long operation time. For the driven KNO, it imposes a more stringent requirement as the physical observable required in the linear response is defined with a couple of quasiorthogonal coherent states. The linear response should be made perturbatively within the steady state subspace constituted by these two basis states, requiring a longer operation time and making even more challenging demands on practical implementation \cite{Wang2023AdvQuantumTechnol}. One method for overcoming such weakness is to amplify the control parameter, which is generally limited by the practical physical systems as the Hamiltonian for stabilizing the steady basis state subspace should be strengthened accordingly. An alternative method is to speed up the operation while maintaining the adiabatic passage of the eigenstate evolution. The latter can be realized by adopting the ``shortcut to adiabaticity" (STA) protocol, where a counterdiabatic Hamiltonian cancels the nonadiabatic deflection of the eigenstate evolution \cite{Wang2018Simulating,Demirplak2003Adiabatic}. We apply such an STA protocol to the established topological model to reduce the operation time for observation of the topological phenomena.   

The paper is organized as follows. In Materials and Methods, we explore the kind of topological phenomena with application of the STA to the established model.In Results and Discussion, we structure the topological Hamiltonian with the driven KNO, whose dynamics for exploration of the topology is controlled within the continuous variable coherent state subspace. In the Conclusion, we summarize.

\begin{figure*}[htbp]
	\centering
{\tiny }	\includegraphics[scale=0.6]{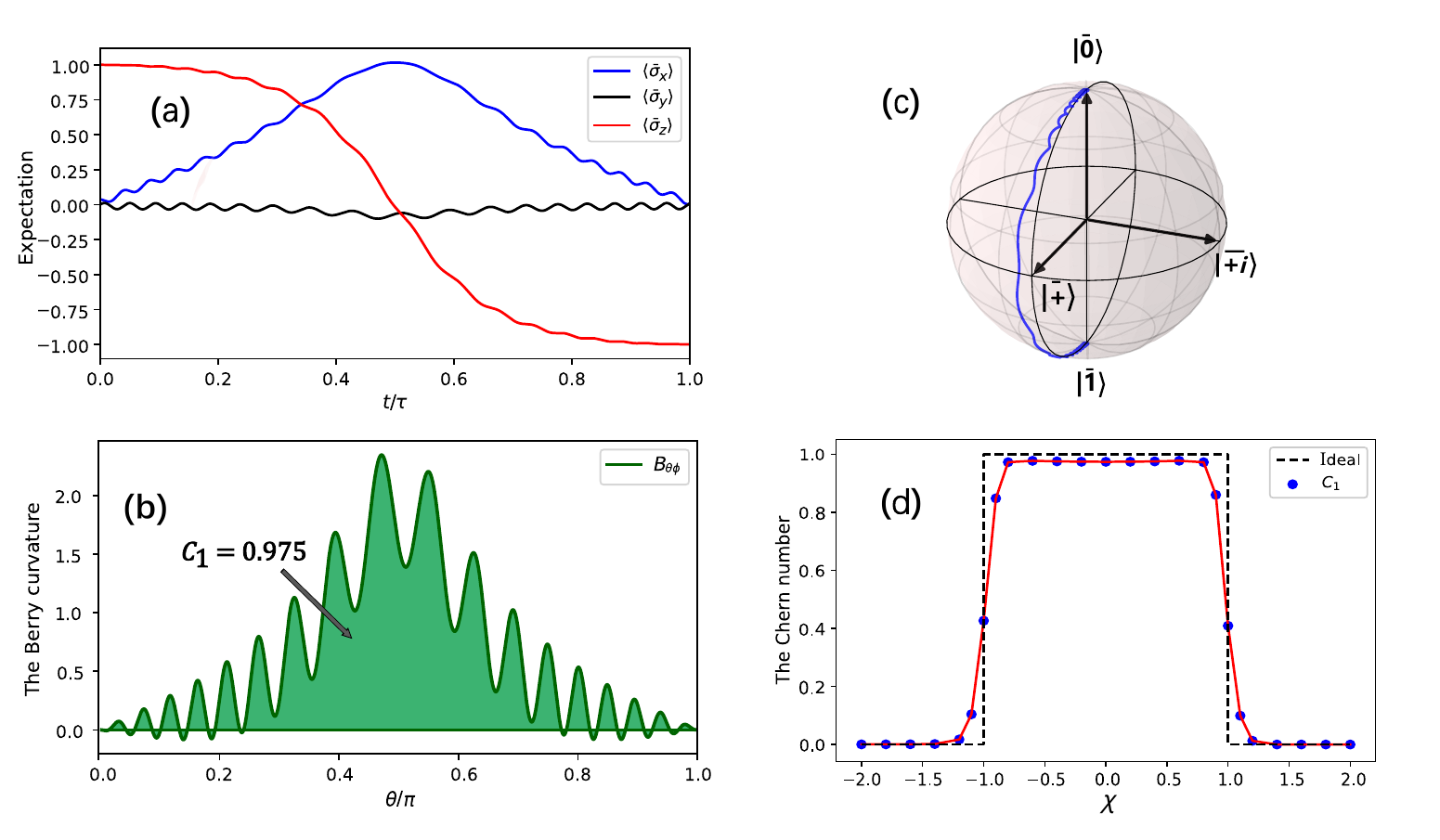}
	\caption{Characterization of the topology of the system. (a) Expectation of the Pauli matrices $\bar{\sigma_j}$ ($j=x,y,z$) versus the operation time $t$ during the linear ramping of the time-dependent polar angle $\theta(t)$, with $\delta_0=0$. (b) The Berry curvature $B_{\theta}$, which is calculated according to Eq. \ref{eq8}, as a function of $\theta$ that ramps linearly with the operation time. The integral of $B_{\theta}$ over the whole parameter $\theta$ gives the first Chern number $C_1=0.975$. The reason of such a nonideal $C_1$ ($<1$) is due to the imperfect ``stabilizer condition". (c) The system state, the blue curve on the Bloch sphere, during the ramping process. (d) The first Chern number $C_1$ as a function of the ratio $\chi$ (the blue dots), which locates the different parameter manifold position. The red curve is the fitting of the calculation result, while the black dashed line is the ideal case for $C_1$.}
	\label{tomo}
\end{figure*}

\section*{Materials and Methods}
Consider the topological Hamiltonian with the driven KNO:
\begin{equation}
	H=H_0+\frac{\Delta_z}{2}H_z+\frac{\Omega}{2}H_x\cos{\phi}+\frac{\Omega}{2}H_y\sin{\phi}. 
\end{equation}
The ``stabilizer role'' $H_0$ makes the system state evolve only within the two-dimensional subspace spanned by $\{|\bar{0}\rangle, |\bar{1}\rangle\}$. The operator $H_j$ ($j = x,y,z$) plays the role of performing single-qubit operations within the continuous variable state subspace $\{|\bar{0}\rangle$, $|\bar{1}\rangle\}$. Confined within such a state subspace, the time-dependent Hamiltonian can be expressed as:
\begin{equation}\label{eq-s2}
	H_{i}=\frac{1}{2}\cdot	\left[   \begin{matrix} {\Delta_z} & {\Omega\cdot e^{-i\phi}} \\ 
		{\Omega\cdot e^{i\phi}} & -{\Delta_z} \end{matrix}   \right].
\end{equation}
The eigenvalues of Eq. \ref{eq-s2} are 
$E_{\pm}=\pm\frac{1}{2}\sqrt{{\Delta_z}^2+\Omega^2}$. The unitary operator $U(t)$ can be defined based on the corresponding eigenstates:
\begin{equation}
	\left[   \begin{matrix} {|\tilde{\psi}_{-}(t)\rangle}  \\ 
		{|\tilde{\psi}_{+}(t)\rangle} \end{matrix}   \right]
	=\left[ \begin{matrix} {-\sin{\frac{\Theta}{2}}} & {\cos{\frac{\Theta}{2}}} \\ 
		{\cos{\frac{\Theta}{2}}} & {\sin{\frac{\Theta}{2}}} \end{matrix}   \right]
	\left[   \begin{matrix} {|\bar{0}\rangle}  \\ 
		{|\bar{1}\rangle} \end{matrix}   \right],
\end{equation}
where $\phi = 0$ is adopted and $\tan{\Theta}=\frac{\Omega}{\Delta_z}$ ($0<\Theta<\pi$). Then the term that causes the deviation from the adiabatic trajectory reads:
\begin{equation}
	-iU(t)\frac{\partial U^{\dag}}{\partial t}=-\frac{\dot{\Theta}}{2}
	\left[ \begin{matrix} {0} & {-i} \\ 
		{i} & {0} \end{matrix}   \right].
\end{equation}

We ignore the ``stabilizer~ role'' $H_0$ and write the general Hamiltonian for the two-level system as:
\begin{eqnarray}
	H_i&=&\bm{R}\cdot\bm{\mathcal{H}}_{\sigma}\\
	&=&X\cdot H_x+Y\cdot H_y+Z\cdot H_z.
\end{eqnarray}
The Berry's phase is given by
\begin{equation}
	\varphi_{Berry}=\iint_S \bm{B}\cdot d\bm{S},
\end{equation}
where $d\bm{S}$ is a vector normal to the surface, and the quantity integrated is the Berry curvature $\bm{B}=\bm{R}/2R^3$, with $\bm{R}$ being the vector of the Hamiltonian $\bm{R} = (X, Y, Z)$. The system states are driven by the time-dependent Hamiltonian and evolve following the adiabatic trajectory. The first Chern number $C_1=\frac{1}{2\pi}\oint_S \bm{B}\cdot d\bm{S}$ can be simplified as:
\begin{eqnarray}
	C_{1q}&=
	&\frac{1}{2\pi}\int_0^{\pi}\int_0^{2\pi}\frac{\bm{R}}{2R^3}\cdot R^2\sin{\theta_q}d\theta_q d\phi\cdot\bm{e_R}\\
	&=&\frac{1}{2}\int_0^{\pi}\sin{\theta_q}\frac{\partial \theta_q}{\partial \theta} d\theta,
\end{eqnarray}
where $\bm{e_R}$ is the unit vector of $\bm{R}$.
\section*{Results and Discussion}
\subsection*{Structuring a topological Hamiltonian using the KNO}
The physical system at hand is the KNO with frequency $\omega_c$ driven by a parametric two-photon drive of a pump frequency $2\omega_c$ \cite{Wielinga1993Quantum,Goto2016Universal,Puri2017Engineering}. In a frame rotating at $\omega_c$ and in the rotating-wave approximation, the Hamiltonian of the system can be described as (setting $\hbar=1$):
\begin{equation}\label{e1}
H_0=-\frac{K}{2}{a^{\dag}}^2a^2+\frac{P}{2}({a^{\dag}}^2+a^2),
\end{equation}
where $a^{\dag}$ and $a$ are the creation and annihilation operators for the KNO, $K$ is the amplitude of the Kerr nonlinearity, and $P$ is the pump amplitude for the two-photon drive. The Hamiltonian of Eq. \ref{e1} possesses two coherent eigenstates $|\pm \sqrt{P/K}\rangle$ with the same eigenvalue $\frac{P^2}{2K}$. 
When the amplitude $P$ is large enough compared to $K$ so that $\langle-\alpha_0|\alpha_0\rangle=e^{-2|\alpha_0|^2}$ is negligible (setting $\alpha_0=\sqrt{P/K}$), the two coherent states $|\alpha_0\rangle$ and $|-\alpha_0\rangle$ are nearly orthogonal to each other, and a pair of basis states $|\bar{0}\rangle\equiv|\alpha_0\rangle$ and $|\bar{1}\rangle\equiv|-\alpha_0\rangle$ can be defined to act as the qubit states. It is then necessary to achieve the basic rotation operators, i.e., Pauli matrices $\bar{\sigma}_{j}$ ($j = x, y, z$), that are defined in the coherent state subspace $\{|\bar{0}\rangle, |\bar{1}\rangle\}$ \cite{Puri2017Engineering}.

To perform the $\bar\sigma_z$ operation on the KNO, we apply a drive field with a pulse-shaped amplitude $1/(2\alpha_0)$, and the Hamiltonian is given by
\begin{equation}\label{eq2}
H_z=(a^{\dag}+a)/2\alpha_0.
\end{equation}
For the `stabilizer' amplitude \cite{Puri2017Engineering}, the system is approximately kept in the subspace $\{|\bar{0}\rangle, |\bar{1}\rangle\}$. Thus the single-photon drive Hamiltonian of Eq. \ref{eq2} realizes the Pauli matrix $\bar\sigma_z$: $\bar I H_z\bar I=\bar\sigma_z$, where the unit matrix is defined as $\bar I=|\bar0\rangle\langle\bar0|+|\bar1\rangle\langle\bar1|$ and $\bar\sigma_z=|\bar0\rangle\langle\bar0|-|\bar1\rangle\langle\bar1|$. For the $\bar{\sigma}_x$ operation, we can consider a specific detuning $\Delta_z $ ($\frac{1}{|2\alpha_0|}e^{2|\alpha_0|^2}$) between the two-photon drive and the resonator, and the corresponding Hamiltonian can be modelled as
\begin{equation}
H_x=\frac{1}{|2\alpha_0|}e^{2|\alpha_0|^2}a^{\dag}a, 
\end{equation}
which projects the number operator in the logical basis $\bar I (H_x)\bar I=e^{2|\alpha_0|^2}\bar{I}-\bar\sigma_x$, where $\bar\sigma_x=|\bar0\rangle\langle\bar1|+|\bar1\rangle\langle\bar0|$. 
The operator $\bar\sigma_y$ can be achieved by a drive field whose Hamiltonian is:
\begin{equation}\label{eq4}
H_y=\frac{-i}{2\alpha_0}e^{2|\alpha_0|^2}(a^{\dag}-a).
\end{equation}
In the subspace spanned by $\{|\bar0\rangle, |\bar1\rangle\}$, the Hamiltonian Eq. \ref{eq4} can be expressed as $\bar IH_y\bar{I}=\bar\sigma_y$, where $\bar\sigma_y=-i|\bar0\rangle\langle\bar1|+i|\bar1\rangle\langle\bar0|$.

We thus model the topological Hamiltonian with the driven KNO as follows:
\begin{equation}\label{topo_H}
H=H_0+\frac{\Delta_z}{2}H_z+\frac{\Omega}{2} H_x\cos{\phi}+\frac{\Omega}{2} H_y\sin{\phi}.
\end{equation}
The related parameters of Eq. \ref{topo_H} can be tailored in such a way that a parameter space with the elliptical manifold is formed:
\begin{equation}\label{eq6}
\Delta_z=\delta_z\cos{\theta(t)}+\delta_0,\quad\Omega=\Omega_0\sin{\theta(t)}.
\end{equation}
Note that the first term, $H_0$, of Eq. \ref{topo_H}, plays the `stabilizer' role and makes the system state evolve only in a two-dimensional subspace spanned by $\{|\bar0\rangle$, $|\bar1\rangle\}$. This can be guaranteed when the strength of $H_0$ is much larger than those of $H_{x,y,z}$.

In our scheme, the system is initialized at $|\bar{0}\rangle =|\sqrt{2}\rangle$ ($\alpha_0=\sqrt{2}$), and during the whole dynamics the time-dependent polar angle is controlled to range from  $\theta(0)=0$ to $\theta(\tau)=\pi$, while the azimuth angle $\phi(t)=0$ is kept, where $t$ [$\in (0,\tau) $] is the operation time. The Berry curvature $B_{\theta\phi}$ of the system is obtained by the linear response of the physical observable $\langle\bar{\sigma}_y\rangle$ to the changing rate of $\theta(t)$: $v_{\theta}=\frac{\partial\theta}{\partial t}$, and can be expressed as \cite{Gritsev2012Dynamical,Roushan2014Observation}:
\begin{equation}\label{eq7}
B_{\theta\phi}=-\frac{\langle\partial_{\phi}H\rangle}{2v_{\theta}}=-\frac{\Omega_0\sin{\theta}}{2v_{\theta}}\langle\bar\sigma_y\rangle.
\end{equation}
Thus the first Chern number can be obtained by integrating $B_{\theta\phi}$ over the whole parameter space $(\theta, \phi)$ \cite{Berry1984Quantal}: $C_1=\frac{1}{2\pi}\int_0^{\pi}d\theta\int_0^{2\pi}d\phi B_{\theta\phi}$. As the Hamiltonian whose parameters, of Eq. \ref{eq6}, is cylindrically symmetric with regard to the $z$ axis, the Berry curvature is a function of $\theta$ solely: $B_{\theta}$, whose integral over $\theta$ directly gives the first Chern number $C_1$:
\begin{equation}\label{eq8}
C_1 = \int_0^{\pi}B_{\theta}d\theta.
\end{equation}

\begin{figure}[htbp!]
	\centering
	\includegraphics[scale=0.5]{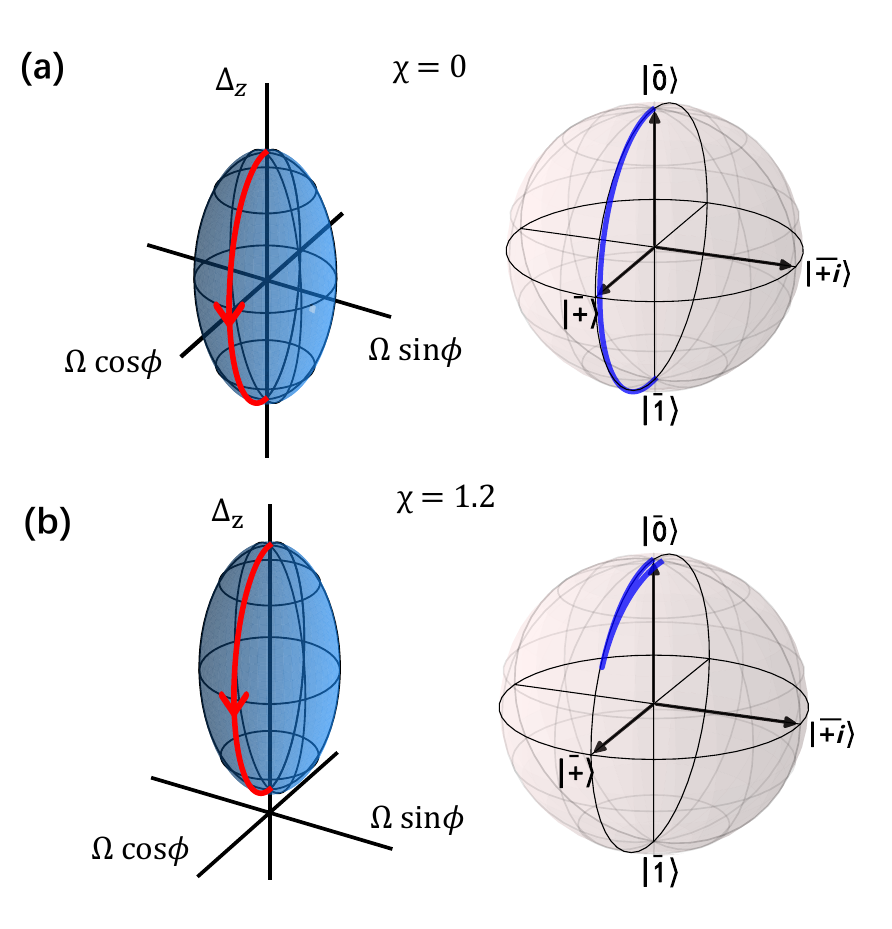}
	\caption{Trajectories of the parameter ramping and of the corresponding motion of the system, for an STA protocol. (a) $\chi=0$. The left shows the blue ellipsoid as the parameter manifold, and the red directional curve represents the ramping path of $\theta$. The right shows the result for the ramping of the system state, which wraps the Bloch sphere, with $\theta \in \left[0, \pi\right] $. (b) The same as (a), but for $\chi=1.2$. The result shows that the motion of the system wraps only part of the Bloch sphere. }
	\label{Fig_bloch}
\end{figure}

Taking into account the ``stabilizer condition", we set $P=2K=1,000$ MHZ, ${\Omega_0=P/(10\cdot e^{2|\alpha_0|^2})}$, $\delta_z=2\Omega_0$, and adopt the linear variation relation of $\theta$ with respect to the operation $t$: $\theta(t)=\pi t/\tau$, with the maximal delay $\tau$ set as $40$ $\mu s$. Fig. \ref{tomo} shows our calculation results for characterization of the topology of the system topology. Fig. \ref{tomo} (a) plots the expectation values $\langle \bar{\sigma}_j \rangle$ ($j=x,y,z$) with respect to the operation time, clearly illustrating the correctness of such physical observables during the linear ramping of $\theta$, as is consistent with the previous case of the discrete variable two-level system \cite{Schroer2014Measuring}. Fig. \ref{tomo} (b) shows the Berry curvature $B_{\theta }$ that is extracted from the linear response of the physical observable $\bar{\sigma}_y$ to $v_{\theta}$. The integral of $B_{\theta}$ over the whole $\theta$ gives the value that is equivalent to the green color area: 0.975, for the parameters as chosen above. Ideal $C_1$ of unity should be obtained if all these parameters are reselected to improve the linear response process \cite{Roushan2014Observation,Schroer2014Measuring}. The motion of the system during the whole ramping of $\theta$, as described by the transient states on the Bloch sphere, is shown in Fig. \ref{tomo} (c) (blue curve). Obviously, the motion trajectory is deflected from that of the ideal ground state evolution (black curve). For $\delta_0 = 0$, the motion state of the system is opposite at the poles: initial state $|\bar{0}\rangle$ and final state $|\bar{1}\rangle$, which is also guaranteed by the expectation $\langle \bar{\sigma}_z \rangle = \pm 1$ for the beginning and end of the ramping process, as shown in Fig. \ref{tomo} (a). The motion in such a case essentially wraps the whole sphere in between. As $\delta_0$ is changed, the motion of the system is quantitatively altered. In Fig. \ref{tomo} (d), the first Chern number $C_1$ is plotted according to the variance of the ratio $\chi \equiv \delta_0/\delta_z$, where $\delta_0$ locates the position of the Hamiltonian parameter manifold of the system. Evidently, the $C_1$ shows two transitions, $C_1=0 \rightarrow 1$ at $\chi =-1$ ($C_1=1 \rightarrow 0$ at $\chi =1$), corresponding to moving the degenerate point ($\Delta_z = \Omega =0$) from outside to inside (contrarily) the parameter manifold. The deviation of the calculated results (the blue dots with its red fitting) from the ideal case (the black dashed line) is due to the established imperfect linear response process based on the chosen parameters. The different $\delta_0$ essentially reflects the different motional dynamics of the system, that is, whether the evolution trajectory wraps (or does not wrap) the whole Bloch sphere. 

Notice that optimizing the ``stabilizer condition" means to decrease the ratio $e^{2|\alpha|^2} \Omega/P$, at the expense of increasing the operation time, thus leading to large decoherence effects \cite{Wang2023AdvQuantumTechnol}. In the next section, an STA protocol is utilized to speed up the operation while keeping it balanced with the ``stabilizer condition".   

\begin{figure}[t]
	\centering
	\includegraphics[scale=0.45]{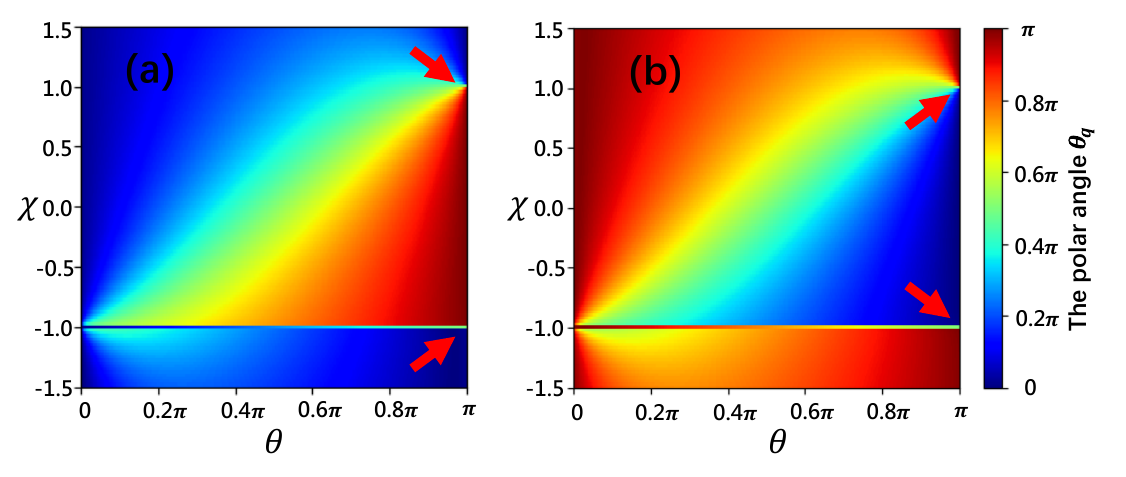}
	\caption{Simulation results of the new polar angle $\theta_q$. (a) For the system initialized as $|\bar0\rangle$, the $\theta_q$ as a function of the ratio $\chi$ and the time-dependent polar angle $\theta(t)$. The red arrows mark the points ($\chi = 1, -1$ at $\theta = \pi$) separating the two topological phases. (b) The same as (a), but with the initial state of the system as $|\bar1\rangle$. The $\theta_q$ changes in contrast to that of (a).}
	\label{theta_q}
\end{figure}

\subsection*{The STA protocol for exploration of topology}
Consider that a system whose state at arbitrary time is described by $|\psi(t)\rangle$, which is dominated by the control Hamiltonian $H(t)$. The transformation to the eigenstate picture of $H(t)$ gives the transformed state $|\tilde{\psi}(t)\rangle=U(t)|\psi(t)\rangle$, where $U(t)$ is the  unitary operator whose rows consist of the eigenvectors of $H(t)$. With such a transformation of $U(t)$, the time-dependent Schr\"{o}dinger equation can be written as \cite{Demirplak2003Adiabatic}:
\begin{equation}
i\frac{\partial|\tilde{\psi}(t)\rangle}{\partial t}=\Big[U(t)H(t)U^{\dag}(t)-iU(t)\frac{\partial U^{\dag}(t)}{\partial t}\Big]|\tilde{\psi}(t)\rangle.
\label{eq_sch}
\end{equation}
The second item on the right side of Eq. \ref{eq_sch} will cause the state evolution to deviate from the adiabatic trajectory. To eliminate the nonadiabatic transition requires an additional counterdiabatic drive field, $H_{cd}=i\frac{\partial U^{\dag}(t)}{\partial t}U(t)$, to be included in its Hamiltonian: $H \rightarrow H + H_{cd}$. With the setting of $\delta_z=\Omega_0$ and $\phi(t)=0$, the STA field $H_{cd}$ can be written as $\dot\Theta\cdot \bar{\sigma}_y/2$ (see Materials and Methods), where $\Theta=\arctan{\frac{\sin{\theta(t)}}{\cos{\theta(t)}+\chi}}$. The total Hamiltonian becomes
\begin{equation}
H_t=H_0+\frac{\Delta_z}{2}H_z+\frac{\Omega}{2}H_x+\frac{\dot{\Theta}}{2}\bar{\sigma}_y.
\label{STA_H}
\end{equation}
Under the setting of $P = 2K = 1,000$ MHZ and $\Omega_0 = \delta_z = 2\pi \times 20$ KHZ, and with the adoption of the time-dependent polar angle $\theta(t)=\frac{\pi}{2}[1-\cos{(\pi t/\tau)}]$ and of the operation delay $\tau=1.5~\mu$s, the trajectory of the ramping of $\theta$ and that of the motion state of the system, starting from initial state $|\bar0\rangle$, for $\chi=0$ and $\chi=1.2$, are calculated as specific illustrations, shown in Fig. \ref{Fig_bloch} (a) and (b), respectively. The figure clearly shows that, for $\chi = 0$, the motion of the system wraps the Bloch sphere accompanying the ramping of $\theta $ from $0$ to $\pi$, whereas for $\chi = 1.2 $ it travels only a part of the path and turns back to the origin. The acceleration effect of the STA protocol is of great importance for exploration of the topology dynamics of the system \cite{Wang2023AdvQuantumTechnol}. 

\begin{figure}[htbp!]
	\centering
	\includegraphics[scale=0.5]{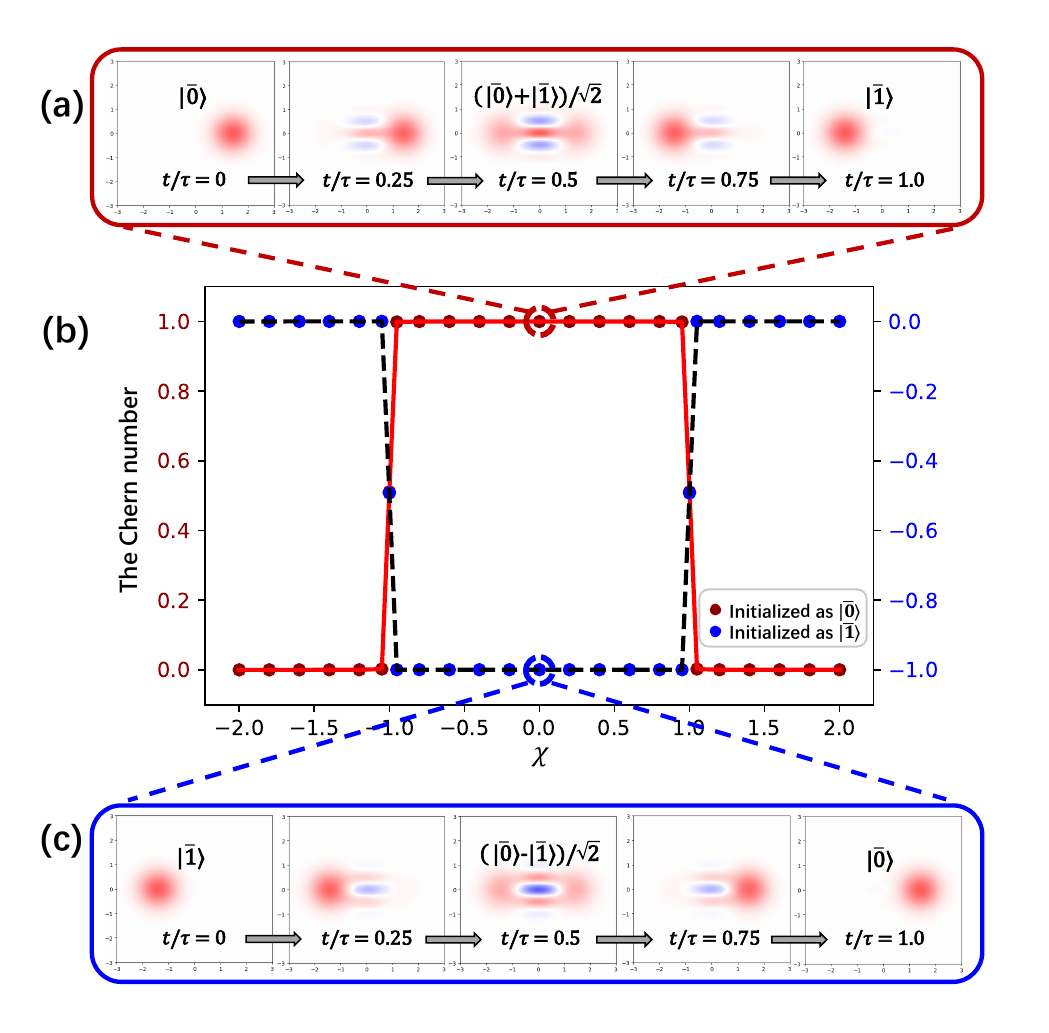}
	\caption{Simulation results for the STA protocol. The Wigner function evolution for the KNO state during the STA (for $\chi = 0$), (a) for initial state $|\bar{0}\rangle$ and (c) for initial state $|\bar{1}\rangle$. (b) The first Chern number $C_{1q}$ as a function of the ratio $\chi$ during the STA. The red dots correspond to initial state $|\bar0\rangle$, and the blue dots correspond to initial state $|\bar1\rangle$. $P=2K=1,000 $ MHZ, $\Omega_0 = \delta_z$ $= 2\pi \times 20$ KHZ, $\theta(t) = \frac{\pi}{2}[1-cos(\pi t/\tau)]$, and $\tau = 1.5$ $\mu s$.}
	\label{Fig_C1}
\end{figure} 

The implementation of the STA protocol means that the measurement of deviation in the physical observable $\langle\bar{\sigma}_y\rangle$ during the adiabatic process has been eliminated, so that the Berry curvature $B_{\theta\phi}$ cannot be extracted by the linear response method through $\frac{\Omega_0\sin{\theta}}{2v_{\theta}}\langle\bar\sigma_y\rangle$. 
However, we can adopt the definition of the new polar angle \cite{Wang2018Simulating}: 
\begin{equation}
\theta_q(t)=\arccos{[\langle\bar{\sigma}_z\rangle/\sqrt{\langle\bar{\sigma}_x\rangle^2+\langle\bar{\sigma}_y\rangle^2+\langle\bar{\sigma}_z\rangle^2}]},
\label{Eq_polar}
\end{equation}
which can be obtained through measurement of the three physical observables $\bar{\sigma}_j$ ($j = x, y, z$) to get their expectations: $\langle \bar{\sigma_j} \rangle$. 

For the system initialized as $|\bar0\rangle$, the simulation results of the $\theta_q(\theta)$ are plotted against the ratio $\chi$ and the time-dependent polar angle $\theta$, as shown in Fig. \ref{theta_q} (a). For the case of $|\chi|<1$, the $\theta_q$ monotonically increases with the increase of $\theta$ to finally reach the almost same value $\pi$ when $\theta$ approaches $\pi$; for $|\chi|>1$, the $\theta_q$ starts with an initial increase but returns to $0$ at $\theta=\pi$. Remarkably, the $\theta_q$ approximately increases linearly with $\theta/2$ at $\chi=1$ or $\chi = -1$, each one at $\theta = \pi$ acts as the critical parameter (marked by the red arrow) separating the two topological phases.  Fig. \ref{theta_q} (b) shows the case for initial state $|\bar1\rangle$. The variation of the $\theta_q$ is just the opposite to that of initial state $|\bar0\rangle$. For $|\chi|<1$, the $\theta_q$ monotonically decreases with the increase of $\theta$ and tends to reach $0$ when $\theta$ arrives at $\pi$;  for $|\chi|>1$, the $\theta_q$ starts with an initial decrease but comes back to $\pi$ at $\theta=\pi$. 

The first Chern number can be calculated through the thus-obtained $\theta_q(t)$ \cite{Roushan2014Observation,Wang2018Simulating}:
\begin{equation}
C_{1q}=\frac{1}{2}\int_0^{\pi}\sin{\theta_q}\cdot\partial_{\theta}\theta_qd\theta,
\end{equation}
where $\partial_{\theta}\theta_q = \frac{\partial \theta_q}{\partial \theta}$. The state describing the motion of the system during the STA process can be tracked through the Wigner function tomography \cite{Wigner1932On,Lutterbach1997Method,Vlastakis2013cat} $W(\alpha) =\frac{2}{\pi}Tr[D_{\alpha}^{\dag}\rho D_{\alpha}P]$, where $D_{\alpha} = e^{\alpha a^{\dag}-\alpha^* a}$ is the oscillator displacement operator, $P = e^{i\pi a^{\dag}a}$ is the photon number parity operator, and $\rho$ is the KNO state density matrix. The numerical simulation results for the STA protocol are shown in Fig. \ref{Fig_C1}. In Fig. \ref{Fig_C1} (b), the $C_{1q}$ is plotted against the ratio $\chi$, for the two cases with the KNO initialized in $|\bar{0}\rangle$ (the red dots) and $|\bar{1}\rangle$ (the blue dots) state. Obviously, for initial state $|\bar{0}\rangle$ ($|\bar{1}\rangle$), the $C_{1q}$ exhibits sharp transitions at both $\chi= -1$: $C_{1q} = 0 \rightarrow C_{1q} = 1$ ($C_{1q} = 1 \rightarrow C_{1q} = 0$), and $\chi =1$: $C_{1q} = 1 \rightarrow C_{1q} = 0$ ($C_{1q} = 0 \rightarrow C_{1q} = 1$), which correspond to the cases of moving the degeneracy from outside to inside (vice versa), and from inside to outside (vice versa), the elliptical manifold, respectively. Fig. \ref{Fig_C1} (a) and (c) plot the Wigner function evolution characterizing the KNO state at the specific moments: $t = (0,0.25,0.5,0.75,1)\tau$, for the cases of initial states $|\bar{0}\rangle$ and $|\bar{1}\rangle$, respectively. This verifies the ideal state evolution confined within each eigenstate assisted by the STA, as illustrated by the Wigner functions uncovering the motion state of the system at specific instant \cite{Haroche2006ExploringQuantum}.

Remarkably, due to the application of the STA, the operation time for extracting the first Chern number is optimized to be largely reduced, which is of great importance from the decoherence point of view \cite{Wang2023AdvQuantumTechnol,Haroche2006ExploringQuantum}.    

\section*{Conclusion}
In summary, we have provided a strategy for the observation of topological transitions in a controlled KNO. The topological invariant first Chern number is characterized by the extraction of the Berry curvature through the linear response of a physical observable to the quench velocity of the system and by integration of it over the whole parameter space of the Hamiltonian of the system. The key to exploring the topology is the linear response induced by deflection from the adiabatic trajectory following the eigenstate evolution. The alternative method is the adoption of the STA to speed up the adiabatic operation. The topological properties are revealed by the introduced new polar angle, which is calculated through the measurement of the three physical observables defined within some of the continuous variable coherent states. Our strategy for the observation of topological properties by the construction of controllable quantum systems is aligned with the concept of quantum simulation first proposed by Feynman \cite{Feynman1982IJTP}, which is of importance for the exploration of possibly inaccessible quantum phenomena. Recent development of versatile superconducting microwave resonator or acoustic-wave resonator design and control techniques provides the potential for the implementation of the strategy \cite{Grimm2020Nature,Marti2024NatPhys}. The proposed strategy can be extended to two or more interacting KNO systems \cite{Grimm2020Nature, Cai2024NatPhys,Zhang2024SCCPM} for the investigation of interaction-involved topological properties in continuous variable systems.

\section*{Acknowledgments}
\subsection*{Funding}
This work was supported by the National Natural Science Foundation of China under Grant Nos. 12274080, 12204105, and 11875108.
\subsection*{Author contributions}
Z.B.Y. conceived and supervised the study. J.L. and S.B.Y. performed the numerical calculations and the plotting. Z.B.Y., F.W., S.B.Y., and J.L. analyzed the data. Z.B.Y. and F.W. co-wrote the paper. All authors contributed to interpretation of the results and helped to improve presentation of the paper.
\subsection*{Competing interests}
The authors declare that they have no competing interests.

\end{document}